\documentclass[11pt]{article}
\usepackage{amssymb,latexsym,amsmath,amsthm}

\newtheorem{theorem}{Theorem}
\makeatletter \@addtoreset{figure}{section}
\def\fps@figure{h, t}
\@addtoreset{table}{bsection}
\def\thetable{\thesection.\@arabic\c@table}
\def\fps@table{h, t}
\newtheorem{proposition}[theorem]{Proposition}

\newfont{\tenbi}{cmbxti10}
\newcommand{\la}{\lambda}

\DeclareMathOperator{\Prym}{Prym}
\DeclareMathOperator{\Jac}{Jac}

\begin{document}

\title{Generic hyperelliptic Prym varieties in a generalized H\'enon--Heiles 
system\footnote{{\small AMS Subject Classification 14H70, 14H40, 70H06, 37J35 }} }

\date{V.Z. Enolski, \\
School of Mathematics, University of Edinburgh, Edinburgh. \\
 On leave from Institute of Magnetism, National Academy of Sciences of
Ukraine, Kiev, {\small Viktor.Enolskiy@ed.ac.uk}, \\ 
Yu.N.Fedorov \\
Department of Mathematics I, Politechnic university of Catalonia, Barcelona, 
{\small Yuri.Fedorov@upc.edu}, \\
A.N.W. Hone, \\ School of Mathematics, Statistics \& Actuarial Science, University of Kent, Canterbury, United Kingdom \\
{\small A.N.W.Hone@kent.ac.uk} }

\maketitle 

\begin{abstract} It is known that the Jacobian of an algebraic curve which is a 2-fold covering of a hyperelliptic curve
ramified at two points contains a hyperelliptic Prym variety.  
Its explicit algebraic description is applied to some of the integrable H\'enon--Heiles systems
with a non-polynomial potential. Namely, we identify the generic complex invariant manifolds of the systems
as a hyperelliptic Prym subvariety of the Jacobian of the spectral curve of the corresponding Lax representation.   
 
The exact discretization of the system is described as a translation on the Prym variety. 
\end{abstract}

\section{Introduction} Many algebraic completely integrable systems possess matrix Lax representations whose spectral curves
admit symmetries (in particular, involutions). The Jacobians of the curves contain Abelian (Prym) subvarieties 
whose open subsets are identified with the complex invariant manifolds of the systems. 

In most cases the Prym subvariety itself is not a Jacobian variety, in particular, due to the fact that  
it is not principally
polarized. In low dimensions the subvariety can be related to the Jacobian of an algebraic curve via an isogeny, and
the latter curve also appears as the curve of separation of variables for the system (see \cite{AvM88, HvM89}, for example). 

On the other hand, as we know from \cite{Mum, Dal}, 
when the spectral curve ${\widetilde C}$ admits an involution $\sigma$ with two fixed points, and it covers a hyperelliptic
curve, say $C$, the corresponding Prym variety becomes the Jacobian of (another) hyperelliptic curve $C'$, and thus
can be referred to as a {\it hyperelliptic Prym variety}. 
Such a situation occurs,
in particular, in finite-dimensional reductions (stationary flows) of the Sawada--Kotera and the Kaup--Kupershmidt 
hierarchies of PDEs (see \cite{Wu}) and the associated integrable dynamical systems, such as the H\'enon--Heiles systems 
described in \cite{Fordy,Fordy2}.

The special case where the two branch points of the covering (the images of 
the two fixed points of the involution $\sigma$) are also related to one another by the hyperelliptic involution 
on $C$ was considered in detail by several authors, see e.g., \cite{fay73, VM_Mum, FVh}. 
In that case, the covering curve $\widetilde C$ is also hyperelliptic, and the equation of the second curve $C'$ representing
the Prym variety is derived in a straightforward way. 
The corresponding applications to integrable systems were considered in 
\cite{FS, FVh, Kuz, Taim1, Taim2}. 

For the general case, $\widetilde C$ is not hyperelliptic, and 
an algorithm for calculating the second curve $C'$ was given only recently in \cite{Levin}. 
In the present paper we apply a modification of the latter method  to describe the complex invariant manifolds of 
certain integrable generalizations of the H\'enon--Heiles
system. To be precise, it is shown that the  invariant manifolds of the cases (i) and (iii) of these
systems are the 2-dimensional Prym subvariety of the Jacobian of a trigonal spectral curve ${\widetilde C}$, 
and the second curve $C'$ is identified with 
the curve associated with the separation of variables found previously in \cite{Cos, Ver_Mus_Conte}.   
Moreover, for an exact discretization (B\"acklund transformation) $\cal B$ of the above systems constructed in \cite{Com_Hone_Mus},
we describe each branch of $\cal B$ as a translation on the Prym variety. 
      
\section{Double cover of a hyperelliptic curve with two branch points}
Consider a hyperelliptic genus $g$ curve $C$: $y^2=f(x)$, where
$f(x)$ is a polynomial of degree $2g+1$ with simple roots. Any 2-fold covering of $C$ ramified at two finite points
$P=(x_P,y_P), Q=(x_Q, y_Q) \in C$ ({\it which are not related to each other by the hyperelliptic involution on $C$})
can be written in the form\footnote{Here and below we identify a curve with its regularization.}
$$
{\widetilde C} \, : \; z^2 = y+ h(x), \quad y^2=f(x),
$$
where $h(x)$ is a polynomial of degree $g+1$ such that
$$
h^2(x)-f(x) = (x-x_P) (x-x_Q) \rho^2(x)
$$
with  $\rho(x)$ being a polynomial\footnote{Here $x_P$ or $x_Q$ may or may not coincide with roots of $\rho(x)$.} 
of degree $g$.
Thus $\widetilde C$ admits the involution $\sigma\, : \; (x,y,z) \mapsto (x,y,-z)$, with fixed points 
$(x_P,y_P,0), (x_Q, y_Q,0) \in {\widetilde C} $.  
Then the genus of $\widetilde C$ is $2g$ and the following was shown by D. Mumford and S. Dalaljan \cite{Mum, Dal}: 
\begin{description}
\item{1)} The Jacobian of $ \widetilde C $ contains two $g$-dimensional Abelian subvarieties: Jac$(C)$ and the 
Prym subvariety $\Prym ({\widetilde C},\sigma)$.
 The former is invariant with respect to the involution $\sigma$ extended to $\Jac(\widetilde C)$, whereas the latter
is anti-invariant.  
\item{2)} Prym$({\widetilde C},\sigma)$ is a principally polarized Abelian variety and, moreover, 
is the Jacobian of a {\it hyperelliptic} curve $C'$.
\end{description}

It was further shown recently by A. Levin \cite{Levin} that the second curve $C'$ can be written explicitly as 
\begin{equation}
w^2 = h(x)+ Z, \quad Z^2=h^2(x)-f (x)\equiv (x-x_P) (x-x_Q) \rho^2(x) ,
\end{equation}
which is equivalent to the plane curve
$$
  [w^2-h(x)]^2 = h^2(x)- f(x) \quad \Longrightarrow \quad w^4-2 h(x)w^2 + f(x) =0.
$$
The latter can be transformed to a standard hyperelliptic form which is given 
in \cite{Levin}.

Note that when the polynomial $f(x)$ is of even degree $2g+2$ and $g$ is odd ($g=1,3,5,7,\dots$),
the above formulas are still valid. However, when $g$ is even, a different result holds, which 
is described as follows. 

\begin{theorem} \begin{description} 
\item{(a)} In the case when the polynomial $f(x)$ has even degree $2g+2$ and $g=2,4,6,\dots$,
any covering $\widetilde C \to C$ ramified at 2 finite points $P=(x_P,y_P), Q=(x_Q, y_Q) \in C$ can be written in the form
\begin{equation} \label{2g}
\left\{ y^2 =f(x), \quad z^2= \frac{y+ h(x)}{x-x_P} \quad \textup{or, equivalently,}
\quad z^2= \frac{y+ h(x)}{x-x_Q}  \right\},
\end{equation}
where, $h(x)$ is of degree 
at most $g+1$ and such that
\begin{equation} \label{rhoo}
 h^2(x)-f(x) = (x-x_P)(x-x_Q) \rho^2(x),
\end{equation}
for some polynomial $\rho(x)$.

\item{(b)} The corresponding Prym variety is isomorphic to the Jacobian of a second genus $g$ hyperelliptic curve
$C'$, which can be written in the form
\begin{gather}
\left\{ Y^2 =h^2(x)- f(x)= (x-x_P)(x-x_Q) \rho^2(x), \right. \notag \\
\left.  w^2= \frac{Y + h(x)}{x-x_P} \quad \textup{or, equivalently,} \quad w^2= \frac{Y + h(x)}{x-x_Q}  \right\}.
\label{2nd_cur}
\end{gather}
The latter is transformed to the standard hyperelliptic form $v^2 =P_{2g+2}(u)$, where $P_{2g+2}$ is a polynomial of degree $2g+2$,
by the birational transformation
\begin{equation} \label{tr_hyp}
 x= \frac{x_Q u^2-x_P}{u^2-1}, \quad  w 
= 
\left(\frac{x_Q-x_P}{u^2-1} \right)^{g/2}\, v ,
\end{equation}
with inverse 
$$
 v = \frac{w}{(x-x_Q)^{g/2} }, \quad   
u= \frac{(x-x_Q) w^2 - h(x)}{ \rho (x)\, (x-x_Q) }.
$$

\item{(c)} In the particular case $g=2$, upon setting 
$\rho(x) = A x^2 + B x + C$, the hyperelliptic form of $C'$ has the following structure
\begin{align}
  v^2 & = \frac{1}{ (x_P-x_Q) ^3 } [\, h(x_Q) u^6 + (x_P-x_Q) \rho(x_Q) u^5 
 - \left[ (x_P- x_Q) h'(x_Q) + 3 h(x_Q)\right] u^4 \notag \\
 & \quad + (x_P-x_Q) \left[ \rho (x_P) + \rho(x_Q) - \rho_2 (x_P-x_Q)^2 \right ] u^3 \notag \\
 & \quad - \left[ (x_P - x_Q) h'(x_P) - 3 h(x_P)\right] u^2
  + (x_Q-x_P) \rho(x_P) u - h(x_P) ] ,  \label{hyp_form}
\end{align}
where $h'(x)$ is the derivative of $h(x)$. 
\end{description}
\end{theorem}

\paragraph{Remark.} Note that the equation \eqref{hyp_form} is symmetric with respect to $x_P, x_Q$ in the sense
that once these two values are interchanged, one obtains the equation of a birationally equivalent curve.
\medskip

\noindent{\it Proof of Theorem 1.} We first notice that the function $y+h(x)$ has poles of order $g+1$ at 
$\infty_1, \infty_2$, the two points  at  infinity on $C$.
Then, in view of \eqref{rhoo}, the function $z^2=(y+ h(x))/(x-x_P)$ has 2 {\it simple} zeros
at some finite points $P,Q$, some double zeros elsewhere, and
poles of even degree $g$ at $\infty_1$ and $\infty_2$.
Thus the regularization of \eqref{2g} is ramified over $C$ at $P,Q$ only.

The proof of the first and second items follows the same lines as that in the paper \cite{Levin} 
for the case of  $f(x)$ having odd degree, and it uses the tower of curves constructed in \cite{Dal}. We only add that 
the equivalence of the two alternative forms of the curves in \eqref{2nd_cur} follows from the relation
\begin{align*}
 \frac{Y + h(x)}{x-x_P} \left(  \frac{Y}{(x-x_Q)\rho(x)} \right)^2 & = \frac{Y + h(x)}{x-x_P} \frac{h^2(x)-f(x)}{[ (x-x_Q)\rho(x) ]^2} \\
& = \frac{Y + h(x)}{x-x_P} \frac{ (x-x_P)(x-x_Q) \rho^2(x)}{[ (x-x_Q)\rho(x) ]^2} =  \frac{Y + h(x)}{x-x_Q} ,
\end{align*}
where we used \eqref{rhoo}. Thus $\frac{Y + h(x)}{x-x_P}$,  $\frac{Y + h(x)}{x-x_Q}$ differ by a square factor,
hence the two forms of
 \eqref{2nd_cur} are transformed to each other by change $w \to w Y/[(x-x_Q)\rho(x)]$. 
A similar argument gives the equivalence of the two forms of \eqref{2g}.

 The substitution \eqref{tr_hyp} was first suggested also in \cite{Levin}; applying it to the second version of
\eqref{2nd_cur}, one obtains
$$
   v^2 = \frac{(x-x_Q) \rho(x) u + h \left( \dfrac{x_Q u^2-x_P}{u^2-1}\right ) }{(x- x_Q)^{g+1}} =
\frac{\dfrac{x_Q-x_P}{1-u^2} \rho\left( \dfrac{x_Q u^2-x_P}{u^2-1}\right ) u+ h\left( \dfrac{x_Q u^2-x_P}{u^2-1}\right ) }
{\dfrac{(x_P-x_Q)^{g+1}}{(1-u^2)^{g+1}}},
$$
which, after simplification, gives a polynomial in $u$ of degree $2g+2$ in the right hand side. In the case $g=2$
this gives the equation \eqref{hyp_form}. $\square$

\paragraph{Remark.} The special case of the 2-fold covering $\widetilde C \to C$ when the branch points
$P=(x_P,y_P), Q=(x_Q, y_Q) \in C$ are
related by the hyperelliptic involution, and $\widetilde C$ is itself hyperelliptic,
had been previously studied in detail in many publications, see e.g., \cite{Mum, VM_Mum, FVh}. 
Namely, let $f(x)$ be of odd order and
$x_P= x_Q=x^*$, where $x^*$ is not a root of $f(x)$. The equation of $\widetilde C$ becomes
$$
   z^2 = x-x^*, \quad y^2 = f(x),
$$
and, assuming $x^*=0$ without loss of generality, one obtains $\widetilde C$: $\{y^2 =f(z^2)\}$.
The involution $\sigma \; : \; (z,y) \mapsto (-z, y)$ commutes with the hyperelliptic involution
$\iota \; : \; (z,y) \to (z, -y)$ and yields the second involution $\tau \; : \;(z,y) \mapsto (-z, -y)$ with two fixed (infinite)
points on $\widetilde C$.
The factor $\widetilde C/\tau$ is the second genus $g$ hyperelliptic curve $C' \; : \{v^2 =u f(u)\}$ whose Jacobian is
isomorphic to $\Prym ({\widetilde C},\sigma)$. In a similar way, $\Prym ({\widetilde C},\tau)$ is isomorphic to $\Jac(C)$. 
Various applications of this special case of the covering $\widetilde C \to C$
to integrable systems were considered, amongst others, in \cite{FS, FVh, Kuz, Taim1, Taim2}.

One should stress that in the general case of the covering $\widetilde C \to C$, when
$x_P\ne x_Q$, there is no second involution on $\widetilde C$, so the latter cannot be a 2-fold covering of the second
curve $C'$. In particular, this means that meromorphic differentials on $\widetilde C \to C$, which are anti-invariant
with respect to $\sigma$, cannot be reduced to differentials on $C'$.   

\subsection{Trigonal genus 4 curve and two curves of  genus 2} 
We now apply the above observations
to the trigonal curve
\begin{gather}
\widetilde{C}: \quad W^3+ a z W^2 +b W z^2 + c W + d z^5+ e z^3 + k z=0, \label{tildec} 
\end{gather}
with parameters $a,b,c,d,e,k \in \mathbb{C}$, 
which has genus 4 and admits the involution $\upsilon: (z,W) \mapsto (-z,-W)$ with
two fixed points ${\widetilde P}=(0,0), {\widetilde Q}=\infty$,  
where $\infty$ is the only point at infinity on ${\widetilde C}$.
Thus the involution induces the covering
$\pi: \widetilde{C} \rightarrow C$ over a hyperelliptic genus two curve $C$.
Namely, upon introducing the new variable $x=z/W$,
which is invariant with respect to $\upsilon$, the equation of $\widetilde C$ is rewritten as
$$
d z^4 x^3 + (1+ax+b x^2 + e x^3) z^2 + k x^3+c x^2 =0 .
$$
Solving it with respect to $z^2$, we get the equation
\begin{gather}
 z^2 = \frac{-h(x)+ \sqrt{f(x)}}{2d x^3}, \quad h(x)=ex^3+bx^2+ax+1, \notag \\
f(x) =\alpha x^6 +\beta x^5 +\gamma x^4+\delta x^3+\varepsilon x^2+2ax+1, \notag \\
\alpha=e^2-4dk,\; \; \beta=2be - 4dc, \;\; \gamma=b^2+2ea,\;\;  \delta=2e+2ab, \;\; \varepsilon=a^2+2b. \label{c}
\end{gather}
Making the birational transformation $(x,z) \to \left(x,\frac {z}{\sqrt{2d}\, x} \right)$,
we finally obtain $\widetilde C$ in the ``canonical'' form \eqref{2g}, that is 
\begin{equation} \label{can_1}
 z^2 = \frac{y -h(x)}{x}, \quad y^2 = f(x)
\end{equation}
Thus $\widetilde C$ is a 2-fold covering of the genus 2 curve $C$: $\{y^2 =f(x) \}$ ramified at
the points 
\begin{gather*}
P=\left( x_P=-\frac{c}{k},\quad y_P=1-\frac{ca}{k}+\frac{c^2b}{k^2}-\frac{c^3e}{k^3} \right), \quad
Q=(x_Q=0,\quad y_Q=1).
\end{gather*}
Note that
\begin{equation} \label{cond_main}
h^2(x)-f(x) = (x-x_P)(x-x_Q) \rho^2(x) = x \cdot (x+ c/k)\cdot (2 \sqrt{dk} x^2)^2 = 4d (kx+c) x^5,
\end{equation}
that is, $\rho(x)= 2 \sqrt{dk} x^2$.

The Prym variety  $\mathrm{Prym} (\widetilde{C}, \upsilon)$ is principally polarized and isomorphic to the Jacobian of
the second genus 2 curve $C'$, which, according to Theorem 1, can be written as  follows:
$$
[ w^2 (x-x_Q)-h(x)]^2= h^2(x)-f(x); 
$$
that is,
\begin{equation} \label{C'3}
C' \; : \quad w^4 (x-x_Q)^2 - 2 w^2 (x-x_Q) h(x)+ f(x)=0.
\end{equation}
with $x_Q=0$ and $h(x), f(x)$ described in \eqref{c}. The explicit hyperelliptic form \eqref{hyp_form} of \eqref{C'3}
will be given for some specific values of the parameters $a,b,c,d,e,k$ in the next section.


\section{Trigonal spectral curve for generalized H\'enon-Heiles systems}
Following an observation of A. Fordy \cite{Fordy}, it is known 
that the generalized H\'enon--Heiles system with the Hamiltonian
$$
 H(p,q) = \frac 12 (p_1^2+p_2^2) + c_1 q_1^2+ c_2 q_2^2 
+ {\mathfrak a} q_1 q_2^2
+ \frac {\mathfrak b}{3} q_1^3 - \frac {\ell^2}{2 q_2^2},
$$
that contains an additional non-polynomial term $\ell^2/(2 q_2^2)$ in the potential, is Liouville integrable for
3 sets of the parameters $c_1, c_2, {\mathfrak a}/{\mathfrak b}$, namely when the latter ratio 
takes the values 
$$
 \frac{{\mathfrak a}}{{\mathfrak b}}\quad  = \quad 
\mathrm{(i)} \, \, 1, \quad \mathrm{(ii)} \, \, 6, \quad \mathrm{(iii)}\, \,  16,  
$$
and $c_1$ and $c_2$ are themselves in a suitable ratio in each case. As was done (for $\ell =0$) in \cite{Ch_Ta_We}, 
these three integrable cases can be isolated using Painlev\'e analysis, while Ziglin's theorem 
can be used to establish non-integrability when the ratio  ${\mathfrak a}/{\mathfrak b}$ does 
not take one of these values \cite{Ito}.

It was shown in \cite{Fordy, Fordy2} that the integrable cases,  numbered  (i), (ii), and (iii) above,
can be identified with appropriate finite-dimensional reductions of the integrable Sawada--Kotera (SK), $KdV_5$ and the
Kaup--Kupershmidt (KK) PDEs, respectively. This property enabled the construction of matrix Lax representations for 
each of these three cases.

For $\ell =0$, the general solution for case (ii) was obtained in \cite{Woj} 
in terms of hyperelliptic quadratures 
by a separation of variables in parabolic coordinates in the $(q_1, q_2)$-plane 
(the generalizations of such separable systems were studied in \cite{EEKL}),
while the equations of motion for the cases (i) and (iii) were integrated in terms of elliptic functions
associated to a pair of different elliptic curves in \cite{Chazy, RGC}.

Henceforth we concentrate on the algebro-geometric description of the cases (i) and (iii) for $\ell \ne 0$.

\subsection{Case (i): Lax pair, spectral curve and B\"acklund transformation} 
The Hamiltonian $H_1(p,q)$ and the additional integral $H_2(p,q)$ for the case (i) can be written in the form
\begin{align}
  H_1 & = \frac 12 (p_1^2+p_2^2) + \frac 16 q_1^3 + \frac 12 q_1 q_2^2 - \frac {\ell^2}{2 q_2^2}, \notag \\
H_2 &= H_{20}^2- \ell^2 \left(\frac{p_1^2}{q_2^2}+ \frac 23 q_1 \right), \qquad
H_{20}= p_1 p_2 + \frac 12 q_1^2 q_2+ \frac 16 q_2^3 . \label{HH12}
\end{align}
The integrals commute with respect to the standard Poisson bracket $\{q_j , p_k\}=\delta_{jk}$ 
on the phase space ${\mathbb R}^4$, and
the generic real invariant manifolds of the system are two-dimensional tori.

From \cite{Com_Hone_Mus, Fordy2}, the Hamilton's equations corresponding to $H_1$ admit the following Lax representation
with spectral parameter $\la\in {\mathbb C}$: 
\begin{gather}
 \dot L(\la) = [ L(\la), N(\la) ], \label{Lax0_HH} \\
L(\la) = \begin{pmatrix}
6\lambda q_{1}  & -\frac{1}{2}(3q_{1}^{2}+q_{2}^{2}) & 9\lambda-3p_{1} \\
9\lambda^{2}+3\lambda p_{1} & -3\lambda q_{1}-q_{2}q_{2} & q_{2}^{2} \\
-\frac{1}{2}\lambda (3q_{1}^{2}+q_{2}^{2}) & 9\lambda^2+ \frac{\ell^2}{q_2^2}-p_{2}^2 &
q_{2}p_{2} -3\lambda q_{1}
\end{pmatrix},  \quad
N(\la)= \begin{pmatrix}
 0 & 1 & 0 \\
0 & 0 & 1 \\
\la  & -q_1 & 0 \end{pmatrix};  \label{LHH}
\end{gather}
the dot stands for the derivative with respect to time $t$.

In addition, the paper \cite{Com_Hone_Mus} presented
a family of exact discretizations (B\"acklund transformations) ${\cal B}:\, (p,q)\to (\tilde p, \tilde q)$
for the case (i) of the generalized H\'enon--Heiles system. The mapping ${\cal B}$ depends on a parameter $\mu\in{\mathbb C}$
and is described by the intertwining relation
\begin{gather}
 \widetilde L(\la) M(\la|\mu) = M(\la | \mu )  L(\la) , \label{LaxHH} \\
M(\la | \mu )= \begin{pmatrix}
\mu -\lambda  & -2Y_{2} & 2Y_{1} \\
2Y_{1}\lambda  & 2Y_{1}Y_{2}-\lambda -\mu  & -2Y_{1}^{2} \\
2(Y_{2}-2Y_{1}^{2})\lambda  & 2Y_{2}(Y_{2}-2Y_{1}^{2})+2Y_{1}(\lambda +\mu )
& \begin{gathered} -2Y_{1}(Y_{2}-2Y_{1}^{2}) \\ -\lambda -\mu \end{gathered}
\end{pmatrix} . \label{MHH}
\end{gather}
Here $L(\la)$ is the same as in \eqref{LHH}, and $\widetilde L(\la)$ depends on  $\tilde p, \tilde q$ in the same way as $L(\la)$
does on $p,q$, whereas $Y_1,Y_2$ are certain functions depending
on the old and new variables $p,q,\tilde p, \tilde q$. Thus the mapping $\cal B$ is {\it implicit}.
The procedure of its evaluation was described in \cite{Com_Hone_Mus} and will also be sketched in the next section.

The spectral curve preserved by the continuous H\'enon--Heiles flow and by the transformation $\cal B$ is
\begin{equation} \label{sp_S}
 {\cal S} \,:\, \det (\eta I - L(\la)) \equiv  \eta^3 - \ell^2 \eta - 729 \la^5 + 162 h_1 \la^3 - 9 h_2 \la =0
\end{equation}
with $h_1, h_2$ denoting fixed values of the constants of motion $H_1, H_2$.

One sees that $\cal S$ is of the type \eqref{tildec} with
$$
a=b=0, \quad c=-\ell^2, \quad d=-729, \quad e=162 h_1, \quad k=-9 h_2,
$$
so it is trigonal of genus 4, and admits the involution
$\upsilon: (\la,\eta) \mapsto (-\la,-\eta)$ with two fixed points $(0,0)$ and $\infty$,
hence it is a 2-fold covering of a genus 2 curve $C={\cal S}/\upsilon$.
It follows that $\cal S$ admits a representation of the form \eqref{can_1} with
$$
 h(x) =162\, h_1 \, x^3+1, \quad \rho(x) = 162 \sqrt{h_2}\, x^2, \quad x_P= - \frac{\ell^2}{9 h_2}, \quad x_Q=0.
$$
Then, from item c) of Theorem 1, we immediately obtain the following.

\begin{proposition} \label{Prym_S} The 4-dimensional Jacobian variety $\Jac({\cal S})$
contains the 2-dimensional Prym variety $\Prym({\cal S},\upsilon)$,
which is isomorphic to the Jacobian of the genus 2 curve\footnote{Here
we omitted a power of the constant factor $x_P-x_Q=x_P$ in front of $v^2$.}
\begin{align}
C' \, : \quad v^2 & = h_2^3(9 u^6 -27 u^4 + 27 u^2- 9) + 2\ell^6 (\sqrt{h_2} u + h_1 ) \notag \\
 & \equiv 9 h_2^3 (u^2-1 )^3 + 2\ell^6 (\sqrt{h_2} u + h_1 ) \, .   \label{2nd_curve}
\end{align}
\end{proposition}

Obviously, the variety $\Jac({\cal S})$ also contains another Abelian subvariety: the Jacobian of the curve
$C=S/\upsilon$. As we will see shortly, the complex two-dimensional invariant manifold of the
generalized H\'enon--Heiles (i) system is an open subset of $\Jac(C')$ and not of $\Jac(C)$.

\subsection{Cases (iii) and (i): separation of variables and the canonical transformation}
In case (iii) the Hamiltonian $K_1$ and the additional integral $K_2$ have the form
\begin{align*}
  K_1 & = \frac 12 (p_1^2+p_2^2) + \frac 43 q_1^3 + \frac 14 q_1 q_2^2 - \frac {2\ell^2}{ q_2^2}, \\
K_2^2 &= K_0^2 -\frac{4\ell^2}{3} q_1 - 8 \ell^2 \frac{p_2^2}{q_2^2}- 4 \frac{\ell^2}{q_2^4}, \\
\text{where} \quad K_0^2 & = p_2^4- \frac{1}{72} p_2^6 - \frac{1}{12} q_1^2 q_2^4 + q_1 p_2^2 q_2^2-\frac{1}{3} p_1 p_2 q_2^3 .
\end{align*}

The paper \cite{Ver_Mus_Conte} provided a non-trivial separation of variables for this system, which led to quadratures
related to a genus 2 hyperelliptic curve ${\cal K}$. Using a  birational
canonical (also non-trivial !)  transformation between the cases (i) and (iii), found
in \cite{En_Sal} for $\ell=0$ and in \cite{Bl_Woj, Com_Hone_Mus} for the general case, these quadratures were
also adopted for a linearization of the case (i) of the H\'enon--Heiles system. Namely, the canonical transformation
implies the following relation between the above integrals $H_1, H_2$ and $K_1, K_2$:
$$
K_1 =H_1, \quad K_2^2 = 4 H_2. 
$$
Then, following  section V of \cite{Ver_Mus_Conte}, for the case (i) the coordinates $q_1, q_2$ read
\begin{equation} \label{qp}
\begin{aligned}
q_1 & = \sqrt{-3} \frac{V_1 -V_2}{U_1-U_2}+ U_1^2+U_1 U_2+ U_2^2 + \frac{3h_2}{\ell^2},  \\
q_2^2 & = -2 \sqrt{-3} (U_1+ U_2) \frac{U_1 V_1 -U_2 V_2}{U_1-U_2}+
2 (U_1+ U_2)^2 \left( U_1^2+ U_2^2+ \frac{9 h_2}{2\ell^2} \right),
\end{aligned}
\end{equation}
where, as above, $h_1, h_2$ are values of the integrals $H_1, H_2$,
and $(U_1, V_1), (U_2, V_2)$ are conjugated separating variables: coordinates of two points on the genus 2 curve
\begin{equation} \label{verh_curve}
{\cal K}\, : \quad
V^2 = P_6(U) = 2 h_1 - \frac 13 \left( U^2+ \frac{3h_2}{\ell^2} \right)^3 - \frac 23 \sqrt{-3 \ell^2}\, U .
\end{equation}
A similar rational expressions for $p_1$ and the product $q_2p_2$ in terms of $(U_1, V_1), (U_2, V_2)$ can be given. 

One can observe that $q_1,q_2^2, p_1, q_2 p_2$ are meromorphic functions on $\Jac({\cal K})$, but $q_2, p_2$ are
not. Next, the invariant tori ${\cal I}_h$ of the H\'enon--Heiles systems  and the systems themselves are invariant 
with respect to the involution $\kappa\, :\,  (q_1,q_2,p_1,p_2) \to (q_1,-q_2,p_1,-p_2)$.  
\medskip

The evolution of $(U_1, V_1), (U_2, V_2)$ with respect to time $t$ is described by the Abel--Jacobi equations
\begin{equation} \label{AJ}
\frac{d U_1}{\sqrt{P_6(U_1)}}  + \frac{d U_2}{\sqrt{P_6(U_2)}} =0, \quad
 \frac{U_1 d U_1}{\sqrt{P_6(U_1)}}  + \frac{U_2 d U_2}{\sqrt{P_6(U_2)}} =dt .
\end{equation}
It follows that the generalized H\'enon--Heiles system (i) is linearized on the Jacobian variety of the curve $\cal K$
and the factor of its generic complex invariant manifold ${\cal I}_h$ by $\kappa$ is an open subset of $\Jac ({\cal K})$. 

In this connection, a natural question is that of how $\cal K$ and the genus 4 spectral curve $\cal S$ are related.
The answer comes out immediately when
one observes that the genus 2 curves \eqref{verh_curve} and \eqref{2nd_curve} are birationally equivalent.
Indeed, setting $U= i \sqrt{3 h_2}u/\ell$,  in \eqref{verh_curve} we get
$$
 V^2 = 2 h_1 - \frac{9 h_2^3}{\ell^6} (1-u^2)^3 + 2 \sqrt{h_2}\, u ,
$$
and the polynomial in $u$ differs from the right hand side of \eqref{2nd_curve} only by the factor $\ell^6$.

Since the transformation between the H\'enon-Heiles cases (i) and (iii) is {\it birational} and, 
according to \cite{Ver_Mus_Conte}, 
the variables $q_1, q_2^2, p_1, p_2^2$ for case (iii) are also meromorphic functions on $\Jac({\cal K})$, in view
of Proposition \ref{Prym_S}, we arrive at the following

\begin{proposition} \label{Prym_HH} For generic values of the constants of motion $h_1, h_2$,
the factorized complex invariant manifold ${\cal I}_h/\kappa$ of the continuous H\'enon--Heiles (i) system, 
as well as of its B\"acklund transformation $\cal B$, is an open subset
of the Prym variety $\Prym({\cal S},\upsilon)$ of the spectral curve $\cal S$.
\end{proposition}

\paragraph{Remark.} The expressions \eqref{qp} jointly with
similar expressions for $p_1, p_2$ in terms of the separation variables 
$(U_1, V_1), (U_2, V_2)$ can be inverted. On the other hand, following known theorems (see e.g., \cite{Fa})
the evolution of $U_j,V_j$ by virtue of the Abel equations \eqref{AJ} can be described by means of a $2\times 2$ matrix
Lax representation with a rational spectral parameter.
All this implies that the generalized H\'enon--Heiles systems also admit such a $2\times 2$ matrix Lax pair.
\medskip

It remains to give a geometric description of the transformation $\cal B$, as an addition/translation on
$\Prym({\cal S},\upsilon)$ and $\Jac({\cal K})$, which will be done in the next section.

\section{Algebraic geometrical description of the mapping $\cal B$}
Following \cite{Com_Hone_Mus}, the mapping $\cal B$ is evaluated as follows.
In view of \eqref{MHH},
\begin{equation} \label{det_M}
\det M(\la |\mu)=-\left( \lambda -\mu \right) \left( \lambda +\mu \right)^{2},
\end{equation}
 and
$M(\mu |\mu)$ has a one-dimensional kernel spanned by $\Phi=(1,Y_1, Y_2)^T$, whereas
the kernel of $M(-\mu |\mu)$ is the two-dimensional space generated by $(0,Y_1, Y_2)^T, (Y_2,\mu,0)^T$. One can write
\begin{equation} \label{M_split}
M(-\mu | \mu)= 2 \begin{pmatrix} 1 \\ -Y_1 \\ -Y_2+2 Y_1^2 \end{pmatrix}
\begin{pmatrix} \mu & - Y_2 & Y_1 \end{pmatrix} .
\end{equation}
By applying both sides of the Lax relation \eqref{LaxHH} to $\Phi$ and setting $\la=\mu$, we see that
$\Phi$ is an eigenvector of $L(\mu)$:
$$
  L(\mu) \Phi = \eta^* \Phi
$$
for a point $(\mu,\eta^*)\in {\cal S}$. Assume that the parameter $\mu$ is generic in the sense that
the covering $(\la,\eta) \to \la$ is not ramified above $\la=\mu$.
Then, fixing $\eta^*$  as one of the three  possible eigenvalues of $L(\mu)$,
we determine the values of $Y_1, Y_2$ in $\Phi=(1,Y_1, Y_2)^T$ in terms of $p_i, q_i$ uniquely,
and, therefore fix a branch of the transformation $\cal B$. It follows that for a generic $\mu$ there are three different
branches.

Explicitly, we have
\begin{equation} \label{Y12}
\begin{aligned}
Y_1 (\mu,\eta^*) & = \frac{2(9^2 \mu^3 - (9p_1^2 + 6 q_1 q_2^2)\mu + q_2^2 \eta^*)}
{ 54 \mu^2 q_1 + 18(\eta^* -q_1p_1 + q_2 p_2) \mu - 6\eta^* p_1- 6 q_2 H_{20}  }, \\
Y_2 (\mu,\eta^*) & = 
\frac{2\eta^2 + 2 q_2 p_2 \eta  - 9(q_1^2- q_2^2) \mu^2 -(6q_1\eta -3q_2^2p_1-9q_1^2p_1+12q_1 q_2p_2)\mu }
{ 54 \mu^2 q_1 + 18(\mu-q_1p_1 + q_2 p_2)- 6\mu p_1- 6 q_2 H_{20} },
\end{aligned}
\end{equation}
with $H_{20}$ being as defined in \eqref{HH12}. 
Substituting these expressions into $M(\la| \mu)$ in \eqref{MHH}, from \eqref{LaxHH} one can evaluate
$\tilde q_i, \tilde p_i$ following the steps indicated in section 5 of \cite{Com_Hone_Mus}.

\paragraph{The mapping $\cal B$ as a translation on $\Prym({\cal S},\upsilon)$ and $\Jac({\cal K})$.}
Consider the eigenvector bundle $ {\mathbb P}^2\to {\cal S} $ associated with the Lax matrix $L(\la)$ in \eqref{LHH},
$$
{\cal S} \ni P=(\la,\eta) \longrightarrow \psi (P) = ( \psi_1(P), \psi_2(P), \psi_3(P))^T \quad \text{such that} \quad
L(\la) \psi(P) = \eta \psi(P) , 
$$
and the equivalence class $\{ {\cal D} \}$ of effective divisors of poles of $\psi (P)$ on the spectral curve $\cal S$.
According to the theory (see e.g., \cite{Dub_Kr_Nov}) for a generic $r\times r$ Lax matrix $L(\la)$,
any divisor $ {\cal D} \in \{ {\cal D} \}$ is given by $n=g+r-1$ points $P_1,\dots, P_n$
on the genus $g$ spectral curve $\cal S$, so in our
case deg ${\cal D}=6$. The equivalence class $\{ {\cal D} \}$ defines a point in $\Jac ({\cal S})$ via the Abel
(Albanese) map 
$$
   {\cal D} \; \mapsto \; \int_{P_0}^{P_1} \bar\omega + \cdots + \int_{P_0}^{P_n} \bar\omega ,  
$$
where $P_0$ is a base point of the map and $\bar\omega$ is a vector of holomorphic differentials forming a basis in
$H^1({\cal S},{\mathbb C})$.  

Let now $\widetilde\psi (P)$ be a section of the eigenvector bundle associated with $\widetilde L(\la)$ and
$\widetilde {\{ {\cal D} \}}$ be the corresponding equivalence class. To describe explicitly the relation between
$ {\cal D}$ and $\widetilde {\cal D}$, and, therefore, the corresponding translation on $\Jac ({\cal S})$,
we follow the standard procedure explained, for example, in \cite{VM_Mum}.

First, as follows from the form of the discrete Lax representation \eqref{LaxHH},
$$
\widetilde\psi (P)= M(\la | \mu)\psi (P).
$$
Hence, ${\cal D}$ and $\widetilde {\cal D}$ can differ by fixed points ${\cal Q}$ on $\cal S$ at which $M(\la,\mu)$
is degenerate
($M(\la | \mu)\psi ({\cal Q})=0$)
or its determinant has a pole. In view of \eqref{det_M}, the only candidates for such points are
\begin{gather*}
 Q_1 = (\mu, \eta_1 ), \quad Q_2 =(\mu, \eta_2), \quad Q_3 =(\mu, \eta_3), \\
 \upsilon(Q_1) =(-\mu, -\eta_1) , \quad \upsilon (Q_2) =(-\mu, -\eta_2), \quad  \upsilon (Q_3)=(-\mu, -\eta_3) ,
\end{gather*}
i.e. the points over $\la=\mu$ and $\la=-\mu$ respectively, and the point at infinity on $\cal S$.

\begin{proposition} \label{Prym_shift} Let us fix a branch of $\cal B$  by choosing $(\mu,\eta^*) = Q_\alpha$,
$\alpha\in \{1,2,3\}$. Then 
\begin{description}
\item{1)} The relation between ${\cal D}$ and $\widetilde {\cal D}$ is
\begin{equation} \label{D_D}
 \widetilde{\cal D} + Q_\alpha +  \upsilon(Q_\beta) + \upsilon(Q_\gamma) \equiv {\cal D} + 3 \infty, \qquad
(\alpha,\beta,\gamma) =(1,2,3).
\end{equation}

\item{2)} The shift divisor
$$
{\cal U}= \widetilde{\cal D}-{\cal D} \equiv Q_\alpha + \upsilon(Q_\beta)+\upsilon(Q_\gamma) - 3 \infty
$$
is equivalent to ${\cal U} \equiv Q_\alpha- \upsilon(Q_\alpha)$ and thus
defines a point on the subvariety  $ \Prym({\cal S},\upsilon) \subset \Jac({\cal S})$. 
\end{description}
\end{proposition}

\paragraph{Corollary.} As follows from item 2) of the above proposition, upon replacing $Q_\alpha=(\mu,\eta^*)$ by 
$\upsilon(Q_\alpha)=(-\mu,-\eta^*)$ the shift $\cal U$ changes sign.
Thus, replacing $M(\la | \mu)$ in the Lax pair \eqref{LaxHH} by $M(\la | -\mu)$ we get the map $\bar{\cal B}$, 
also having 3 branches, which are inverse to those of ${\cal B}$. 

In the special case $\mu=0$ one has $Q_1=(0,0), Q_2 = (0,\ell), Q_3=\upsilon(Q_2)=(0,-\ell)$, 
so the corresponding branches of ${\cal B}$ are the identity map, 
and the two others that are inverse to each other. In the latter case expressions \eqref{Y12} simplify to
$$
  Y_1 =- \frac{q_2^2 \ell}{ q_2 H_{20}+ p_1 \ell }, \qquad Y_2= \frac{\ell (q_2 p_2+ \ell)}{ q_2 H_{20}+ p_1 \ell }; 
$$
however, the explicit formulae for the transformation $(p,q) \to (\tilde p, \tilde q)$ are still too cumbersome to show
here. 

\medskip

\noindent {\it Proof of Proposition} \ref{Prym_shift}.
1) Let $M^*(\la | \mu)$ denote the matrix $M(\la | \mu)$ with the values of $Y_1, Y_2$ chosen above. Then we have
\begin{equation}
\begin{aligned}
 M^*(\mu | \mu) \, \psi(Q_\alpha) & =0, \quad M^*(\mu | \mu)\, \psi(Q_\beta) \ne 0, \quad 
\quad M^*(\mu | \mu)\, \psi(Q_\gamma) \ne 0, \\
  M^*(-\mu | \mu) \,\psi(\upsilon(Q_\alpha))& \ne 0, \quad M^*(-\mu | \mu)\, \psi(\upsilon(Q_\beta)) = 0, \quad
  \; M^*(-\mu | \mu) \, \psi(\upsilon(Q_\gamma)) = 0. 
\end{aligned} \label{-mu}
\end{equation}
Indeed, $\psi(Q_\alpha)$ is proportional to $(1,Y_1, Y_2)^T$, the kernel of $M^*(\mu | \mu)$.
The latter is one-dimensional, which implies the first line in \eqref{-mu}.

Next, by construction, the vectors $\psi(\upsilon( Q_\beta)), \psi(\upsilon(Q_\gamma))$ can be written explicitly as
$$
 (1, Y_1(-\mu,-\eta_\beta), Y_2(-\mu,-\eta_\beta))^T, \quad \text{respectively} \quad
 (1, Y_1(-\mu,-\eta_\gamma), Y_2(-\mu,-\eta_\gamma))^T,
$$
where
$Y_1(-\mu,-\eta_\beta), Y_2(-\mu,-\eta_\beta))$ are evaluated as in \eqref{Y12}. 
Due to \eqref{M_split},
$M^*(-\mu,\mu) \, \psi(\upsilon(Q_\beta))$ is proportional to 
\begin{gather*}
\Pi_\beta (1, -Y_1, -Y_2 +2Y_2^2)^T , \\ 
\Pi_\beta = \langle \, ( \mu, - Y_2(\mu,\eta^*), Y_1(\mu,\eta^*))^T, (1, Y_1(-\mu,-\eta_\beta), 
Y_2(-\mu,-\eta_\beta))^T \rangle . 
\end{gather*}
Explicitly, up to a constant factor, the above scalar product reads
\begin{align}
\Pi_\beta & =  q_2^2 \left[ 729 \mu^5 - 27 (3p_1^2+3p_2^2 + q_1^3 +3 q_1 q_2)\mu^3 + 9 H_{20}^2 \mu \right] \notag \\
 & \quad + ((\eta^*)^2 + \eta^* \eta_\beta+ \eta_\beta^2) [9^2 \mu^3 -3(2 q_1 q_2^2 + 3p_1^2) \mu ]
+ \eta^* \eta_\beta (\eta^* + \eta_\beta ) q_2^2 . \label{Pib}
\end{align}
Due to the form of the spectral curve $\cal S$ in \eqref{sp_S},
$$
  \eta_\alpha + \eta_\beta+ \eta_\gamma =0, \quad
\eta_\alpha \eta_\beta + \eta_\gamma\eta_\alpha + \eta_\beta \eta_\gamma = - \ell^2 ,
$$
which implies
$$
(\eta^*)^2 + \eta^* \eta_\beta+ \eta_\beta^2 = \ell^2, \quad
\eta^* \eta_\beta (\eta^* + \eta_\beta ) = -\eta^* \eta_\beta \eta_\gamma = - \det L(\mu).
$$
Then the right hand side of \eqref{Pib} can be written as
$$
   q_2^2 \left[ 729 \mu^5 - 162 h_1 \mu^3 + 9 h_2 \mu -  \det L(\mu) \right],
$$
which is zero in view of the equation of $\cal S$. Hence $M^*(-\mu,\mu) \, \psi(\upsilon(Q_\beta))=0$. 
 
The same argument shows that $M^*(-\mu,\mu) \psi(\upsilon(Q_\gamma))=0$. Since
$\psi(\upsilon(Q_\beta)), \psi(\upsilon(Q_\gamma))$ span the whole kernel of $M^*(-\mu,\mu)$, we conclude that
$M^*(-\mu,\mu) \psi(\upsilon(Q_\alpha))\ne 0$.
Thus, the inverse operator $(M^*(\la,\mu))^{-1}$ acting on $\widetilde\psi(P)$, $P\in {\cal S}$
produces poles of $\psi(P)$ only at $Q_\alpha, \upsilon(Q_\beta), \upsilon(Q_\gamma)$.
Therefore, to get $\cal D$, one must add the above points to $\widetilde{\cal D}$.
On the other hand, $\widetilde{\cal D}$ is obtained from $\cal D$ by adding a multiple of $\infty\in {\cal S}$.
Since the degree of $\widetilde{\cal D}-{\cal D}$ must be zero, we obtain the relation \eqref{D_D}.
\medskip

2) Adding to $\cal U$ the divisor of the meromorphic function $1/(\la+\mu)$, that is 
$$3\infty- \upsilon(Q_1)- \upsilon(Q_2)- \upsilon(Q_3),
$$ 
 we get $Q_\alpha- \upsilon(Q_\alpha)$, which is obviously
anti-invariant
with respect to the involution $\upsilon$. Thus $\cal U$ defines a point on the Prym variety,
by the definition of the latter. $\square$

\paragraph{Remark.} Since $\Prym({\cal S},\upsilon)$ is identified with the Jacobian of the curve $\cal K$ associated
with the separation of variables, it is natural to describe the shift $Q_\alpha- \upsilon(Q_\alpha)$ on Prym explicitly 
as
an equivalence class $\{ {\cal P}_1 - {\cal P}_2 \}$, where ${\cal P}_1, {\cal P}_2$ are effective divisors of 
equal degree on $\cal K$. This can be made by considering Jacobians of the curves which form the corresponding tower (tree) of curves
given in \cite{Dal}. We want to avoid this specific analysis in the present paper, so 
an explicit derivation of ${\cal P}_1, {\cal P}_2$ will be made in another publication in a broader context 
of hyperelliptic Prym varieties.          

\section*{Acknowledgments}
The authors are grateful to Aaron Levin for valuable comments concerning Theorem 1 and to Harry Braden for stimulating
discussions. 
The work of V.E was supported by the School of Mathematics, University of Edinburgh, with the certificate of sponsorship 
C5E7V94128U. Yu.F acknowledges the support of the MICIIN grants MTM2012-31714 and MTM2012-37070.

\end{document}